# Continuous Tuning the Magnitude and Direction of Spin-Orbit Torque Using Bilayer Heavy Metals

*Pan He[1a], Xuepeng Qiu[1a], Vanessa L. Zhang[2], Yang Wu[1], Meng Hau Kuok[2], and Hyunsoo Yang[1]\**

[1]Department of Electrical and Computer Engineering, National University of Singapore, 4 Engineering Drive 3, 117576, Singapore
[2]Department of Physics, National University of Singapore, 2 Science Drive 3, 117551, Singapore

\*E-mail: eleyang@nus.edu.sg
[a]Authors with equal contribution

Spin-orbit torques (SOTs) have opened a new path to switch the magnetization in perpendicularly magnetized films and are of great interest due to their potential applications in novel data storage technology, such as the magnetic random access memory (MRAM). The effective manipulation of SOT has thus become an important step towards these applications. Here, current induced spin-orbit effective fields and magnetization switching are investigated in Pt/Ta/CoFeB/MgO structures with bilayer heavy metals. With a fixed thickness (1 nm) of the Ta layer, the magnitude and sign of current induced spin-orbit effective fields can be continuously tuned by changing the Pt layer thickness, consistent with the current induced magnetization switching data. The ratio of longitudinal to transverse spin-orbit effective fields is found to be determined by the Ta/CoFeB interface and can be continuously tuned by changing the Pt layer thickness. The Dzyaloshinskii-Moriya interaction (DMI) is found to be weak and shows an insignificant variation with the Pt thickness. The results demonstrate an effective method to tune SOTs utilizing bilayer heavy metals without affecting the DMI, a desirable feature which will be useful for the design of SOT-based devices.

1. Introduction

Spin-orbit torques (SOTs) induced by in-plane currents have been intensively studied in heavy metal (HM)/ferromagnet (FM)/capping multilayers, in which a HM is adjacent to a thin FM capped by an oxide[1-13] or metal.[9,14,15] SOTs have been recently demonstrated as an alternative approach to switch the magnetization efficiently,[1-7] drive fast chirality dependent domain wall motion,[8-10] as well as excite high-frequency magnetization oscillations.[11-13] The spin Hall effect (SHE)[16-18] and/or Rashba effect[1,19,20] arising from spin-orbit coupling (SOC), have been discussed as the mechanisms giving rise to SOTs, although their relative contributions are currently under debate.[1,3,7] On the other hand, SOTs were also discovered in a bare FM layer and Co/Pd multilayers.[21-23] Tuning the magnitude and direction of SOT, which is an important step towards the applications of SOT-based devices, has been demonstrated using different HM thicknesses[5] and oxygen manipulation,[7] respectively. However, the ability to simultaneously and continuously tune the magnitude and direction of SOT is still lacking.

The HM layer with strong SOC can produce a vertical ($z$ axis) spin current $J_S$ with spin polarization along the $y$ axis, when an in-plane charge current $J_C$ is applied along the $x$ axis (see **Figure 1**a). HMs with opposite spin Hall angles, such as Pt (positive) and Ta (negative), generate reversed spin currents, thus giving rise to opposite SOTs in the FM layer.[1-5,8,24] SHE induced SOTs have been generally understood to be a source of longitudinal SOT effective field ($H_L$) along the $M \times y$ direction, and proposed as the main mechanism to switch the magnetization and drive domain wall motion.[2-14] On the other hand, the Rashba effect, due to the interface inversion asymmetry of the FM layer, is thought to induce a transverse effective magnetic field ($H_T$) and also acts as another source of SOTs.[1,19,20,25] Further studies are still needed to comprehensively understand the sources of $H_L$ and $H_T$ as debates on their origins still exist.[4,5,24] Moreover, the roles of $H_L$ and $H_T$ in the current induced magnetization switching remain unclear. The ratio of $H_T$ to $H_L$ has been found to be critical for current induced switching[26] and thus its effective tuning is important for applications. In order to obtain the direction and amplitude of $H_T$ and $H_L$, harmonic Hall voltage measurements have been widely used,[4,5,7,23,24] along with spin-torque ferromagnetic resonance.[2,3,27]

Recently, the Dzyaloshinskii-Moriya interaction (DMI) has been found to play an important role in the current induced magnetization switching through chiral domain wall nucleation and motion by SOT, [8-10,28-31] while its detrimental effect on switching has been reported in some other scenarios.[26,32] DMI is also found to be responsible for the formation of magnetic skyrmions,[33,34] which serve as bits to store information in future memory and logic devices. A HM capping layer on top of a FM has been proposed in the multilayer systems to tune the interfacial DMI and thus domain wall motion, which also changes the SOT.[9,14,15] If SOTs can be effectively tuned in the multilayer systems without changing the DMI, it will provide a versatile and novel way to design SOT-based devices.

In order to elucidate the above questions, we have conducted measurements on the Pt/Ta/CoFeB/MgO multilayer structure with a HM underlayer of Pt/Ta. The thickness of the Pt layer is changed from 0 to 8 nm, while the thickness of Ta layer adjacent to the CoFeB layer is kept at 1 nm, so that the same Ta/CoFeB interface is maintained with changing the thickness of the bottom Pt layer. Due to opposite spin Hall angles in Pt and Ta, as schematically shown in Figure 1a, the magnitude and sign of the net SOT can be varied by altering the Pt thickness. Using harmonic Hall voltage measurements, we show that both the magnitude and sign of $H_L$ and $H_T$ change simultaneously with increasing the Pt thickness. Furthermore, the current induced switching efficiency is found to be consistent with the measured spin-orbit effective fields. However, the DMI is found to be weak and quite independent of the Pt thickness. Therefore, the SOT and DMI in this bilayer HM system can be tuned separately.

## 2. Results and Discussion

The multilayer films composed of Pt ($t$)/Ta (1)/Co$_{40}$Fe$_{40}$B$_{20}$ (0.9)/MgO (1.2)/SiO$_2$ (3) (numbers in parentheses are nominal thicknesses in nm) were patterned into Hall bars with dimensions of 10 μm in width as shown in Figure 1b. The easy axis of the CoFeB magnetization lies in the out-of-plane direction due to the interfacial perpendicular magnetic anisotropy,[35] as demonstrated by the anomalous Hall loops in Figure 1c. In order to quantify the current induced SOT effective fields, we have measured the harmonic Hall voltages in two schemes,[4,5,7,23,24] namely the longitudinal and transverse geometry. During the measurements, a sinusoidal current $I_{ac}$ with a low frequency ($f$ = 13.7 Hz) and proper amplitude is applied to the devices, and the first and second harmonic Hall voltage signals ($V_f$ and $V_{2f}$, respectively) are measured simultaneously by two lock-in amplifiers while sweeping the magnetic field $H$ applied nearly in plane with $\theta_H$ = 86° (with respect to the $z$ axis). $V_f$ is determined by the equilibrium position of magnetization, while $V_{2f}$ is due to the a.c. oscillation of the magnetization.[4,24] The applied sinusoidal current generates a.c. SOT effective fields, which lead to small oscillations of the magnetization around its equilibrium position and thus give rise to the second harmonic Hall signal. Therefore, we can extract the SOT effective field by analyzing the second harmonic Hall signal. In the longitudinal scheme, both the a.c. current and in-plane component of $H$ are applied along the $x$-direction, whereas, for the transverse scheme, the in-plane component of $H$ is applied along the $y$-direction perpendicular to the current flow.

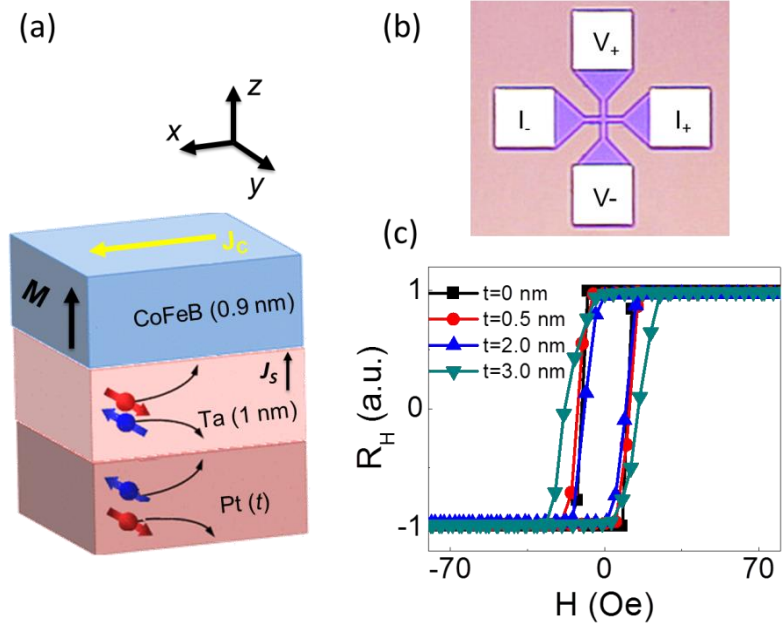

**Figure 1.** a) Schematic of the Pt/Ta/CoFeB multilayer film, and spin currents in each HM layer. b) Typical device image by optical microscope with electrical connection information for Hall measurements. c) Normalized Hall resistance ($R_H$) as a function of out-of-plane field (H) for devices with various Pt thicknesses.

**Figure 2**a,b shows the harmonic Hall voltage signal measured with $I_{ac}$ = 2 mA (peak value) for the device without Pt ($t = 0$) in the longitudinal (∥) and transverse (⊥) schemes, respectively. In the longitudinal geometry, $V_{2f}$ (red squares) shows a negative and positive peak at the positive and negative magnetic field regions, respectively. In contrast, for the transverse geometry, $V_{2f}$ shows a negative peak for both polarities of magnetic field. $V_f$ (black curves) is almost the same in both geometries. The results are consistent with previous studies, in which Ta is used as the HM layer.[4,24] Figure 2c,d presents the harmonic signal measured with $I_{ac}$ = 5 mA for the device with a Pt thickness $t = 3$ nm in the longitudinal and transverse schemes, respectively. There are sign changes in $V_{2f}$ in both geometries as compared to that in Figure 2a,b, whereas $V_f$ has the same sign for both devices. The sign reversal of $V_{2f}$ indicates that the direction of spin-orbit effective fields (both $H_L$ and $H_T$) is reversed, when the 3 nm thick Pt is inserted underneath Ta. As the interface between the CoFeB and Ta layer is the same in both samples, the observed sign change cannot be attributed to any interface effect. Rather, the observed sign reversal of both $H_L$ and $H_T$ can be mainly attributed to the change of SHE induced SOT caused by the insertion of an additional Pt layer. As schematically shown in Figure 1a, Pt and Ta generate spin currents with opposite spin polarizations via SHE, thus giving rise to opposite SOTs in the FM layer. Since the 1 nm thickness of Ta is smaller than its spin diffusion length,[36,37] the spin current from the Pt layer can diffuse through the Ta layer to generate torques in the FM layer.

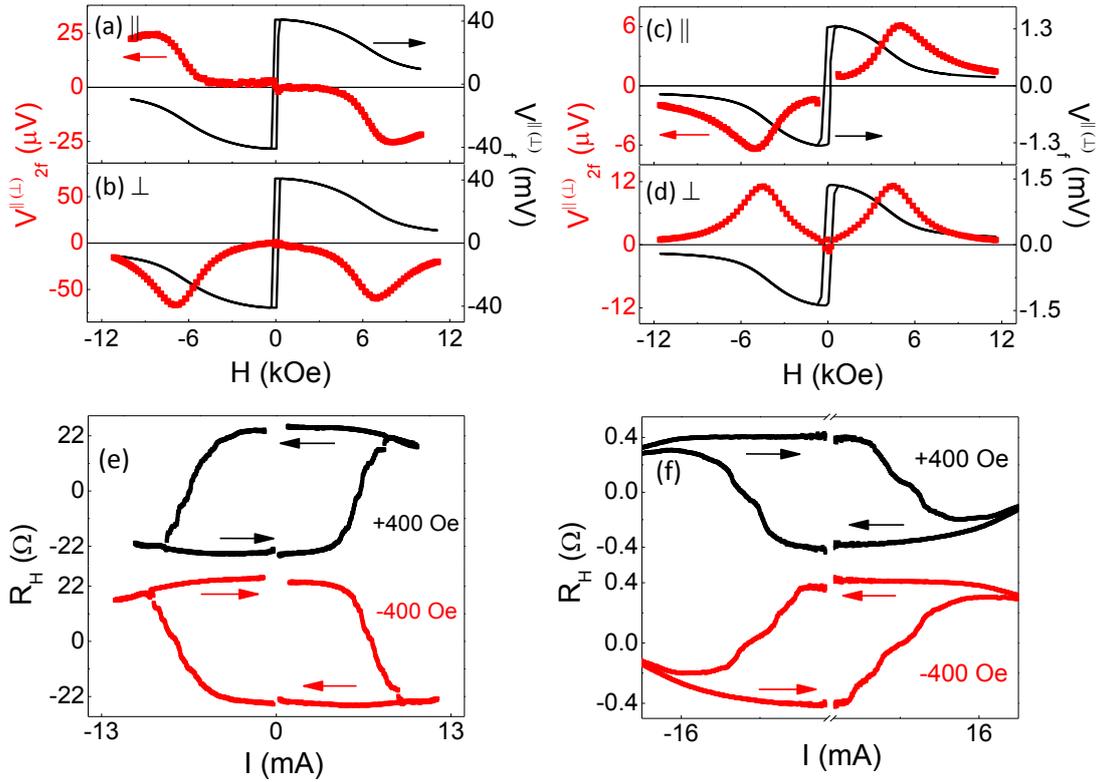

**Figure 2.** First and second harmonic ($V_f$ and $V_{2f}$) Hall loops measured in longitudinal (a,c) and transverse (b,d) schemes with Pt thickness of 0 (a,b) and 3 nm (c,d). e,f) Hall resistance $R_H$ as a function of the amplitude of pulse currents with Pt thickness of 0 (e) and 3 nm (f) at both positive (400 Oe) and negative (-400 Oe) assist field.

It is also expected that the variation of spin-orbit effective fields has an influence on the current induced magnetization switching. Figure 2e,f shows the current induced switching results for the devices with the Pt thickness of $t = 0$ and 3 nm, respectively. During the measurements, an in-plane assist field of 400 Oe is applied either along (+x) or opposite (-x) to the current direction. A clear current induced magnetization switching is observed in both devices, as confirmed by the switching of anomalous Hall signal upon sweeping the in-plane current pulse. Moreover, the switching direction reverses between the two devices ($t = 0$ and 3 nm). Under a positive magnetic field and current, the device without Pt prefers the magnetization along the z-direction ($M_z > 0$), while the other device favors the magnetization along the -z-direction ($M_z < 0$). Furthermore, the DMI is found zero for both devices, as demonstrated below (in Fig. 4) by the Brillouin light scattering (BLS) measurements.[38] Therefore, the reversal of the switching direction observed in Figure 2e,f can be attributed to the reversed spin-orbit effective fields, when a 3 nm thick Pt layer is added under the Ta layer. The sign change of spin-orbit effective fields indicates that the non-local SHE induced SOTs from the 3 nm Pt layer is dominant over that of Ta.[39]

To better understand the effect of Pt in the bilayer HM system, a systematic Pt thickness dependence study has been carried out. **Figure 3**a,b shows the normalized longitudinal $\overline{V}_{2f}^{//}$ and transverse $\overline{V}_{2f}^{\perp}$ second harmonic Hall signal per unit average current density $J_{ave}$, respectively, as a function of $\cos\theta$ with various Pt thicknesses. $\theta$ is the polar angle (with respect to the z axis) of the magnetization defined in a spherical coordinate system with its range from 0° to 86° ($\cos\theta > 0$), and 94° to 180° ($\cos\theta < 0$) in our experiment, and the normalized second harmonic Hall signal per unit average current density is defined as $\overline{V}_{2f}^{//(\perp)} = V_{2f}^{//(\perp)} \cdot H_K / (V_{AHE} \cdot J_{ave})$. The saturated anomalous Hall voltage, $V_{AHE}$ was obtained by measuring the out-of-plane Hall loop. The perpendicular anisotropy field, $H_K$ can be obtained by fitting the first harmonic signal $V_f = V_{AHE}\cos\theta$ using the equilibrium equation of the magnetization, $\sin 2\theta / \sin(\theta_H - \theta) = 2H / H_K$,[40] where $\theta_H$ is the polar angle of the applied magnetic field, which is 86° (for a positive magnetic field) in our measurements. A typical example of the $H_K$ fitting is shown in *Supporting Information Figure S1* and $H_K$ ranges from 0.9 to 7.4 kOe in our samples (*see Supporting Information Figure S2*). The analytical expression of second harmonic Hall signal as a function of a.c. spin-orbit effective field amplitude $\Delta H_{L(T)}$ is given in **Equation** 1 and 2. The detailed derivation process of Equation 1 and 2 is provided in the *Supporting Information*.

$$V_{2f}^{//} = \frac{\sin\theta}{\sin 2\theta \cdot \cot(\theta_H - \theta) + 2\cos 2\theta} \cdot \frac{V_{AHE}\Delta H_L}{H_K} + \frac{-\sin(\theta_H - \theta)\sin\theta}{\cos\theta_H \sin\theta + \sin(\theta_H - \theta)} \cdot \frac{V_{PHE}\Delta H_T}{H_K}, \quad (1)$$

$$V_{2f}^{\perp} = \frac{1}{2\cot(\theta_H - \theta) + 4\cot 2\theta} \cdot \frac{V_{AHE}\Delta H_T}{H_K} + \frac{-\sin(\theta_H - \theta)\sin\theta}{\sin\theta_H \cos\theta} \cdot \frac{V_{PHE}\Delta H_L}{H_K}. \quad (2)$$

Equations 1 and 2 show that the magnitude of second harmonic Hall voltage is not only related to $\Delta H_{L(T)}$, but also to $\theta_H$, $\theta$, $V_{AHE}$, $V_{PHE}$, and $H_K$. Thus, the normalized second harmonic Hall signal $\overline{V}_{2f}^{//(\perp)}$, which removes the influence of $V_{AHE}$ and $H_K$, is more straightforward for evaluating $\Delta H_{L(T)}$. When the planar Hall effect (PHE) is much smaller than the anomalous Hall effect (AHE), the second terms on the right side of Equations 1 and 2 are negligible. The normalized second harmonic Hall signal $\overline{V}_{2f}^{//(\perp)}$ can then be used as a simple estimation of the longitudinal (transverse) spin-orbit effective field $\Delta H_{L(T)}$ (per unit current density) according to the above equations.

Figure 3a,b shows that both $\overline{V}_{2f}^{//}$ and $\overline{V}_{2f}^{\perp}$ change gradually from negative to positive values with increasing Pt thickness at positive magnetic fields, indicating the gradual variation of spin-orbit effective fields $\Delta H_L$ and $\Delta H_T$, respectively. The measured ratio $\alpha$ of PHE to AHE ($\alpha = V_{PHE}/V_{AHE}$) increases gradually with Pt thickness from 2.4% ($t = 0$) to 16% ($t = 3$ nm), showing PHE is much smaller than AHE. The exact values of $\Delta H_L$ and $\Delta H_T$ can be obtained by fitting the second harmonic Hall signal, taking into consideration contributions from both the AHE and PHE. Here we define $u(\theta) = \frac{\sin\theta}{\sin 2\theta \cdot \cot(\theta_H - \theta) + 2\cos 2\theta}$, $v(\theta) = \frac{-\sin(\theta_H - \theta)\sin\theta}{\cos\theta_H \sin\theta + \sin(\theta_H - \theta)}$. Equation 1 can be simplified as

$$V_{2f}^{\parallel} = u(\theta) \cdot \frac{V_{AHE} \Delta H_L}{H_K} + \alpha v(\theta) \cdot \frac{V_{AHE} \Delta H_T}{H_K},$$ thus we can obtain the normalized longitudinal second harmonic Hall signal at a given current as,

$$\frac{V_{2f}^{\parallel} \cdot H_K}{V_{AHE}} = u(\theta) \cdot \Delta H_L + \alpha v(\theta) \cdot \Delta H_T. \quad (3)$$

We also define $x(\theta) = \frac{1}{2\cot(\theta_H - \theta) + 4\cot 2\theta}$, $y(\theta) = \frac{-\sin(\theta_H - \theta)\sin\theta}{\sin\theta_H \cos\theta}$. Equation 2 can be simplified as $V_{2f}^{\perp} = x(\theta) \cdot \frac{V_{AHE} \Delta H_T}{H_K} + \alpha y(\theta) \cdot \frac{V_{AHE} \Delta H_L}{H_K}$, thus we can obtain the normalized transverse second harmonic Hall signal as,

$$\frac{V_{2f}^{\perp} \cdot H_K}{V_{AHE}} = x(\theta) \cdot \Delta H_T + \alpha y(\theta) \cdot \Delta H_L. \quad (4)$$

We can extract $\Delta H_{L(T)}$ by solving two simultaneous Equation 3 and 4. A typical example of the $\Delta H_{L(T)}$ fitting is shown in Supporting Information Figure S3. $\Delta H_{L(T)}$ normally changes with the angle $\theta$ (*see Supporting Information Figure S4*).[4,24]

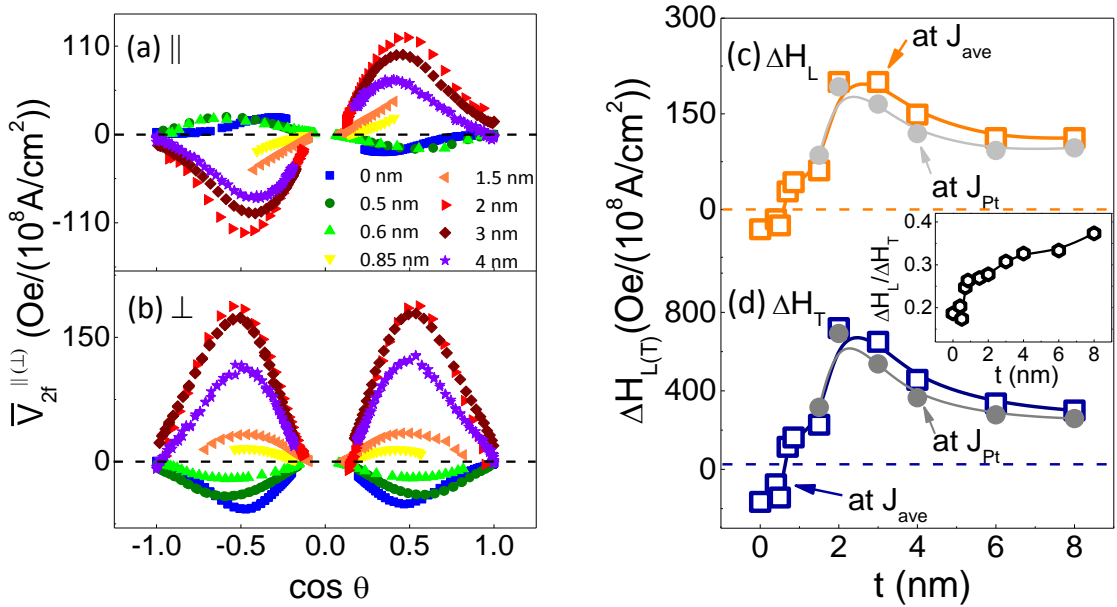

**Figure 3.** a,b) Normalized second harmonic Hall signal at longitudinal a) and transverse b) geometries for devices with different Pt thicknesses. c,d) The fitted longitudinal $\Delta H_L$ (c) and transverse $\Delta H_T$ (d) spin-orbit effective fields (at $+M_z$) at per unit $J_{ave}$ (squares) and $J_{Pt}$ (circles) as a function of Pt thickness (*t*). $\Delta H_L / \Delta H_T$ as a function of Pt thickness (*t*) is shown in the inset.

As shown in Figure 3c,d, the extracted $\Delta H_L$ and $\Delta H_T$ at $+M_z$ at per unit $J_{ave}$ (squares) first increase gradually from negative to positive simultaneously with increasing the Pt thickness and reaches a maximum at $t = 2$ nm, and then, decreases gradually at the relatively thick Pt regime, consistent with the results in Figure 3a,b. Since the resistivity of different metallic layer is different, the current density changes from layer to layer. For simplicity, we presented the data in Figure 3 based on $J_{ave}$. To explain the variation of $\Delta H_L$ and $\Delta H_T$ with $t$, the current distribution in the metallic layers needs to be considered. For detailed discussion about the current density and current distribution, we performed resistivity measurements. The resistivity of Ta (1 nm) and CoFeB (0.9 nm) layers in our multilayer stack is estimated to be 192 and 95 μΩ·cm, respectively. In addition, the resistivity $\rho_{Pt}$ of the Pt layer is found to increase dramatically with decreasing the Pt thickness (*see Supporting Information Figure S5*), which is a well-known phenomenon due to strong diffusive scattering at the Pt surface.[41] The relative current density and current distribution in metallic layers can be obtained (*see Supporting Information Table S1 and S2*). The ratio of distributed currents in Pt to that in Ta increases dramatically with increasing the Pt thickness (*see Supporting Information Table S2, Column 5*). When the Pt thickness is above 2 nm, majority of currents are distributed in the Pt layer. As a result, the spin torque generated from Pt as compared with that from the Ta layer becomes dominant with increasing the Pt thickness. This explains a gradual change of SOT effective fields from negative to positive.

As reported recently, the spin Hall conductivity $\sigma_{SH}$ in the Pt layer is a constant independent of the resistivity, while the spin Hall angle $\theta_{SH} = (2e/\hbar)\sigma_{SH}\rho$ would change proportionally with the resistivity ($\rho$), and the spin diffusion length in Pt has been reported to scale linearly with $1/\rho$.[41,42] The resistivity of Pt layer decreases by ~ 3 times when its thickness increases from 2 to 8 nm (*see Supporting Information Figure S5*). The spin Hall angle of the Pt layer in our samples would therefore decrease accordingly with increasing the Pt thickness. This can explain why $\Delta H_L$ and $\Delta H_T$ in Figure 3c,d decrease when the Pt thickness is above 2 nm. The ratio $J_{Pt}/J_{ave}$ between the current density in Pt and the average current density changes little when the Pt thickness is above 2 nm (*see Supporting Information Table S1*). Therefore, when $\Delta H_L$ and $\Delta H_T$ are at per unit $J_{Pt}$ (circles in Figure 3c,d), there is a similar decreasing trend above 2 nm. The gradual decrease of SOT effective fields (at per unit $J_{Pt}$) at a relatively thick Pt regime has been also reported,[41] which is consistent with our results even though a Pt/Ta bilayer is used in our case.

The observed change of $\Delta H_L$ and $\Delta H_T$ with Pt thickness further indicates that the diffusive spin current generated in the Pt layer plays an important role in our structure. Interestingly, $\Delta H_L$ and $\Delta H_T$ can be tuned close to zero simultaneously at a Pt thickness ~ 0.7 nm. The vanishing $\Delta H_L$ and $\Delta H_T$ may suggest that spin currents from Ta and Pt cancel each other, giving rise to a null SOT to the FM layer. The present results suggest $\Delta H_L$ and $\Delta H_T$ have the same dominant origin and the contributions of SOTs from both HMs should be taken into consideration.[8,9,43] The ratio of $\Delta H_L$ to $\Delta H_T$ is shown in the inset of Figure 3c,d, which increases gradually by two times with Pt thickness (*t*). The ratio of $\Delta H_L$ to $\Delta H_T$ has been discussed as an important parameter in current induced magnetization switching without an assist field[26] and its tunability present here in the bilayer HM structure is an important step towards application. $\Delta H_T$ has been found to be larger than $\Delta H_L$ in Ta based SOT structure before,[5,23,24] which is the case for the sample without Pt (*t* = 0) studied here. However, $\Delta H_T$ has been found to be smaller than $\Delta H_L$ in Pt based SOT structures.[3,4,8,24,25] The gradual increase of $\Delta H_L/\Delta H_T$ with the Pt thickness indicates that the spin current from Pt is playing an increasingly important role with increasing the Pt thickness. Nevertheless, $\Delta H_T$ is found to be larger than $\Delta H_L$ in the Pt/Ta bilayer HM structure even when *t* = 8 nm. The results suggest it is the Ta/CoFeB interface that determines the relative magnitude of $\Delta H_L$ and $\Delta H_T$ and the thickness of Pt underneath Ta/CoFeB interface plays a minor role.

The effect of Pt thickness on current induced magnetization switching has also been systematically studied. The SOT switching efficiency, defined as $H_K/J_c$,[7,24,25,44] is shown in **Figure 4**a as a function of Pt thickness ($t$). $J_c$ is the critical average current density, when the magnetization changes from $+M_z$ to $-M_z$ at the positive assist field. With increasing Pt thickness, the SOT switching efficiency decreases gradually first to zero at $t = 0.7$ nm, where no current induced switching is observed, and then reverses sign and increases followed by a saturation behavior above 2 nm. The variation trend of the SOT switching efficiency with Pt thickness coincides with that of SOT effective fields as shown in Figure 3, indicating a close correlation between the SOT effective fields and the SOT switching efficiency. On the other hand, the switching efficiency presented here in the Pt ($t = 2$ nm)/Ta (1 nm)/CoFeB (0.9 nm) bilayer HM structure is even larger than that of the Pt (2 nm)/CoFeB (0.8 nm) structure.[7] It may indicate that the Ta/CoFeB interface has a better capability in current induced magnetization switching compared with that of the Pt/CoFeB interface. The Pt layer provides the dominant spin current in the Pt (t ≥ 2 nm)/Ta (1 nm)/CoFeB (0.9 nm) multilayer structure. The spin currents generated in the HM layer transport across the HM/FM interface to generate torques in the FM layer. The HM/FM interface plays an essential role in determining the strength and characteristic of the resultant torques,[45] as it determines the spin mixing conductance and interfacial spin scattering, and both of them can influence the interfacial spin transparency and $\Delta H_L/\Delta H_T$ ratio. Moreover, the HM/FM interface determines the DMI of the multilayer system. Here, the Ta/CoFeB interface is found to have a much smaller DMI constant as compared with that of Pt/CoFeB interface. All these factors (spin transparency, $\Delta H_L/\Delta H_T$ ratio and DMI) influence the current induced magnetization switching process.

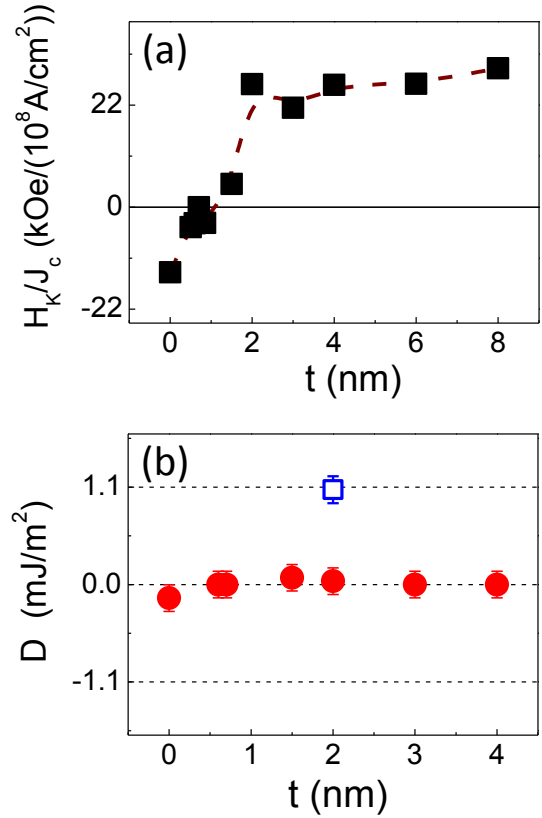

**Figure 4.** a) Anisotropy field ($H_K$) divided by switching current density ($J_c$), defined as the switching from high to low $R_H$ at positive assist field, as a function of Pt thickness ($t$). b) DMI constant ($D$) for samples with different Pt thicknesses (red circles), and a control sample of Si/MgO (2 nm)/Pt (2 nm)/CoFeB (0.8 nm)/MgO (2 nm)/SiO$_2$ (3 nm) shown as a blue square.

BLS measurements on the bilayer HM system were conducted to ascertain the strength of the interfacial DMI.[38,46] BLS spectra (see *Supporting Information Figure S6*) were recorded in the 180°-backscattering geometry using a 6-pass Fabry-Perot interferometer. The results shown in Figure 4b indicate zero DMI in the bilayer HM samples (red circles) similar to a recent report on the Ta/CoFeB interface,[10] in contrast to the reference sample (blue square) with Pt (2 nm)/CoFeB (0.8 nm) interface which has a significant DMI constant, $D$ of 1.1 mJ/m$^2$.[46] It indicates that the spin-orbit coupling property of the Ta/CoFeB interface is not changed by the Pt layer underneath them over a large range of Pt thickness (t = 0 to 8 nm). Therefore, our results demonstrate that SOTs can be tuned by changing the thickness of HM$_1$ (Pt) without affecting the DMI in HM$_1$/HM$_2$/FM/oxide multilayers with bilayer HM, like Pt/Ta/CoFeB/MgO studied here. We also demonstrate a continuous tuning of the magnitude and direction of SOT by changing the $HM_2$ (Ta) thickness in the same bilayer HM structure (see *Supporting Information Figure S7*).

3. Conclusion

In summary, we have demonstrated that SOTs in samples with bilayer heavy metal Pt/Ta can be continuously tuned from positive to negative by varying the thickness of the bottom Pt layer. In particular, the longitudinal and transverse spin-orbit effective fields change simultaneously both in sign and amplitude, providing further insight into their dominant origin. The magnitude of spin-orbit fields is found to be consistent with the efficiency of current induced magnetization switching. Moreover, the ratio of $\Delta H_L$ to $\Delta H_T$, found to be mainly determined by the Ta/CoFeB interface, can be continuously tuned in the bilayer HM structure by changing the Pt thickness from 0 to 8 nm. BLS measurements reveal a negligible change in DMI for such a large variation in Pt thickness. The present experimental results suggest a versatile way to tune SOT without affecting DMI, which will be helpful for designing SOT devices.

4. Experimental Section

The multilayer films composed of Pt (*t*)/Ta (1)/$Co_{40}Fe_{40}B_{20}$ (0.9)/MgO (1.2)/$SiO_2$ (3) (numbers in parentheses are nominal thicknesses in nm) were deposited on thermally oxidized Si substrates at room temperature by ultrahigh vacuum AJA magnetron sputtering (base pressure $< 2\times10^{-9}$ Torr). The deposition rates for Pt, Ta, CoFeB, MgO and $SiO_2$ were 0.50, 0.28, 0.11, 0.03, and 0.08 Å/s, respectively. The film thickness (*t*) of Pt is varied from 0 to 8 nm. No annealing treatment was given to the samples. Before film deposition, the substrates were coated with a layer of positive photoresist (PFI) and the photoresist were patterned into 10-μm-wide Hall bars by photolithography and developing, which was used as a mask for the film deposition. During the deposition, the films are naturally patterned into Hall bars with dimensions of 10 μm in width as shown in Figure 1b. Acetone was used to lift off the photoresist.

Hall bars devices were wire-bonded to the sample holder and installed in a physical property measurement system (Quantum Design) for transport measurements. We performed the measurements of a.c harmonic Hall voltage and current-induced magnetization switching. For the harmonic Hall voltage loop measurement, a Keithley 6221 current source and two Stanford Research SR830 lock-in amplifiers were used. During the measurements, a constant amplitude sinusoidal current with a frequency of 13.7 Hz is applied to the devices, and the in-phase first harmonic $V_f$ and out of phase (90° off) second harmonic $V_{2f}$ Hall voltage signal were measured simultaneously by two lock-in amplifiers while sweeping the magnetic field $H$ applied nearly in plane with $\theta_H = 86°$ (with respect to the $z$ axis). The magnetic field was generated by a Helmholtz coil driven by a Kepco power supply. For the current-induced magnetization switching measurement, a combination of Keithley 6221 current source and 2182A nanovoltmeter were used. During the measurement, a pulsed d.c. current with a duration of 200 μs was applied to the devices and the Hall voltage was measured simultaneously. Meanwhile, an in-plane assist field was applied either along or opposite to the current direction. An interval of 0.1 s was used for the pulsed d.c. current to reduce the accumulated Joule heating effect. All measurements were carried out at room temperature.

BLS measurements were performed in the 180° backscattering geometry and in *ps* polarization using the 514.5 nm radiation of an argon-ion laser and a six-pass tandem Fabry-Perot interferometer. The in-plane magnetic field $H$ was applied perpendicular to the incident plane of light, corresponding to the Damon-Eschbach (DE) geometry. The counter-propagating spin waves with the same wave vector magnitude, -$k$ and +$k$, were simultaneously recorded in a single BLS spectrum as Stokes and anti-Stokes peaks respectively. The frequency difference between the Stokes and anti-Stokes peaks, $\Delta f$, is induced by interfacial DMI and thus provides a direct measurement of the DMI constant, $D =$ $\pi M_s \Delta f/(2\gamma k)$, where $M_s$ is the saturation magnetization and $\gamma$ is gyromagnetic ratio of the ferromagnetic material.

**Supporting Information**
Supporting Information is available.

Acknowledgements

We thank Dr. Ping Yang for XRD measurements. This research was supported by the National Research Foundation (NRF), Prime Minister's Office, Singapore, under its Competitive Research Programme (CRP Award No. NRF CRP12-2013-01), and the Ministry of Education, Singapore (Grant No. R144-000-340-112). H.Y. is a member of the Singapore Spintronics Consortium (SG-SPIN). P. He and X. Qiu contributed equally to this work.